# A data-supported history of bioinformatics tools

Clément Levin[1], Emeric Dynomant[1,2,3], Bruno J Gonzalez[4], Laurent Mouchard[5], David Landsman[6], Eivind Hovig[7,8,9], Kristian Vlahovicek[10]*

[1]omicX, Seine Innopolis, 72 rue de la Republique, 76140 Le-Petit-Quevilly, France.
[2]Normandie Univ, INSA Rouen, LITIS, 76000 Rouen, France.
[3]Département d'Informatique et d'Information Médicales, D2IM, CHU de Rouen, France.
[4]Normandie Univ, UNIROUEN, Inserm U1245 and Rouen University Hospital, Normandy Center for Genomic and Personalized Medicine, Rouen, France.
[5]Normandie Univ, UNIROUEN, LITIS, 76000 Rouen, France.
[6]Computational Biology Branch, National Center for Biotechnology Information, NLM, NIH, USA.
[7]Department of Informatics, University of Oslo, Oslo, Norway.
[8]Department of Tumor Biology, Institute for Cancer Research, The Norwegian Radium Hospital/Oslo University Hospital, Oslo, Norway.
[9]Department of Cancer Genetics and Informatics, Oslo University Hospital, Oslo, Norway.
[10]Bioinformatics Group, Division of Molecular Biology, Department of Biology, Faculty of Science, University of Zagreb, 10000, Zagreb, Croatia.
*Corresponding author: Kristian Vlahovicek, kristian@bioinfo.hr.

**Abstract**

Since the advent of next-generation sequencing in the early 2000s, the volume of bioinformatics software tools and databases has exploded and continues to grow rapidly. Documenting this evolution on a global and time-dependent scale is a challenging task, limited by the scarcity of comprehensive tool repositories. We collected data from over ~23,000 references classified in the OMICtools database, spanning the last 26 years of bioinformatics to present a data-supported snapshot of bioinformatics software tool evolution and the current status, to shed light on future directions and opportunities in this field. The present review explores new aspects of computational biology, including country partnerships, trends in technologies and area of development, research and development (R&D) investments and coding languages. This is the most comprehensive systematic overview of the field to date and provides the community with insights and knowledge on the direction of the development and evolution of bioinformatics software tools, highlighting the increasing complexity of analysis.

## Introduction

Bioinformatics tools and databases constitute an integral component of the current research process in life and biomedical sciences. Since the introduction of information technology in biological research, a plethora of computational tools and databases have followed, contributing to major breakthroughs in the field. One such ground-breaking example is the development of Edman protein sequencing[1] which generated a new form of data at the time, a protein sequence, and fostered the development of the Needleman-Wunsch protein sequence alignment algorithm[2]. Today, several more steps in the development of bioinformatics have been taken, with e.g. high-throughput sequencing (HTS) experiments requiring highly sophisticated and specific pipelines of analysis tools[3,4]. In the past decade, the notion of "biological data" has shifted in magnitude, from sets of hundreds to sets of millions and even billions of entities[5,6]. This exponential increase in the volume of biological data has stimulated the development of an ever-increasing number of bioinformatics tools. Moreover, biological data have shifted in complexity, evolving from a gene-centric perspective to the systems level, and most tasks now involve computer-assisted reduction of large collected raw datasets for conversion into usable knowledge.

A major challenge that today's researchers face is to find the most appropriate and up-to-date tools and resources for their analyses. Online repositories and databases that centralize, reference, classify, and provide access to, available tools represent indispensable aids to the scientific community[7]. Moreover, these repositories represent a focused opportunity to mine for and exploit a wealth of existing information.

Given the dramatic rate at which the field is evolving, still without any signs of decrease, unravelling of the evolutionary patterns of bioinformatics tools and databases is a complex challenge as we generate an ever-increasing amount of data and knowledge on how to store, process, analyze and visualize biological data. While a variety of specialized reviews focusing on software tools for specific areas (e.g. protein structure prediction, secretome analysis, binding site identification, etc.) have been published, to our knowledge, the global perspective and state-of-the-art of the field of "omics" bioinformatics tools is lacking. In this historical perspective, using a comprehensive bioinformatics tool repository, we discuss the early days of bioinformatics through to today and provide an illustrated snapshot of the evolution of bioinformatics in recent decades.

**The interlaced roots of bioinformatics**

The term "bioinformatics" is now widely recognized as an entire field that encompasses biology, medicine, computer science, mathematics, statistics, and information technology. Mainstream applications and concepts in bioinformatics include high-throughput sequencing, "big data" and the internet. The origins of bioinformatics can be set in the late 1950s[8,9]. Since the publication of the DNA structure by Watson and Crick[10], every major breakthrough in the study of genes and proteins has involved informatics technology. As the power and computational capabilities of machines have expanded, so too has our comprehension of the complexity of biology[11].

One of the earliest applications in biology for computers was the calculation of crystallographic structures of proteins, after John Kendrew solved the three-dimensional structure of myoglobin[12]. The same year, a computer was used for the first time to calculate and estimate genetic linkage[13]. The field of biochemistry rapidly followed on with the use of computers, driven by the accumulation of protein sequences after Frederick Sanger determined the sequence of bovine insulin in the early 1950s[14]. Margaret Oakley Dayhoff was the first to compile protein sequences in a database, published in a book format in the 1960s as the Atlas of Protein Sequence and Structure[15], and to develop a computer-assisted sequence alignment and comparison program[16,17]. In the meantime, the nascent field of molecular evolution was revolutionized by the calculation power of computers to generate phylogenic trees[18,19].

By 1970, biologists had successfully developed and implemented computer programs for the study of protein structure, function, and evolution. After 1965, when the first gene sequences were available, it was straightforward for researchers to apply their programs to the biology of nucleic acids[20,21]. The same year, a new comparison method was successfully applied to both protein and nucleic acid sequences[22]. Likewise, novel computer programs were developed for RNA structure prediction[23], pairing schemes of RNA molecules[24], DNA conformation[25], and DNA sequence data handling (storage, edition, translation, comparison, etc.)[26]. In 1977, two new DNA sequencing methods were published[27,28], which greatly extended the amount of nucleic acid sequences to be analyzed.

In parallel with the growing application of computers in biology, the field of informatics itself was profoundly transformed with the creation of the ARPANET in 1969, predecessor of the internet and the World Wide Web that came much later in 1992[29]. Progressively, computers became more accessible for the average researcher. In 1981, IBM launched its first Personal Computer, and just three years later it was used to write a sequence analysis program, rendering large and expensive computers obsolete[30].

By the 1980s, bioinformatics was already at the crossroads of biology, biomedicine, computer science, medicine, statistics, mathematics and physics. It became mainstream in the late 1980s for the management and analysis of sequence data, protein structure determination, function prediction and phylogeny[9]. The progressive accumulation of data stimulated the creation of the first computer-based databases for gene and protein sequences. In the mid-80s, the GenBank and EMBL sequence libraries were launched, which, in the absence of a global interconnected network, at the time were stored and shared on floppy disks or local networks[31,32]. Sequence alignment programs soon became overwhelmed with thousands of sequences to be searched, aligned, and compared. To address this, the FASTA and BLAST database search algorithms were developed and are still widely used[33,34].

**Bioinformatics in the genomics era**

Bioinformatics entered a new era in the 1990s with the advent of genomics, a term used to describe the scientific discipline of mapping, sequencing, and analyzing whole genomes to understand organismal biology in a systems-level approach. The first generations of automated sequencing machines were introduced in 1986 by Applied Bioscience and used the Sanger sequencing method. This revolutionary technology paved the way to the long-awaited whole-genome sequencing of prokaryotic and eukaryotic model organism[35,36], and the completion of the Human Genome Project in the early 2000s[37,38]. This in turn opened the door to a new field of biology, systems biology, dedicated to the study of interactions and networks within biological systems as a whole[39].

In the past two decades, the success of the Human Genome Project and the high demand for low-cost sequencing has fostered the development of "next-generation" sequencing (now referred to as "high-throughput" sequencing), which not only applies to genome sequencing, but also to transcriptome profiling (RNA-sequencing), DNA-protein interactions (ChIP-sequencing), epigenome characterization, evaluation of DNA methylation (BS-seq) and much more[40]. While bioinformatics initially primarily revolved around sequence analysis, genomics and proteomics, today this domain has expanded to encompass the handling of many types of biological data, including text-mining data, imaging, mass spectrometry, flow-cytometry data, etc.

Recent bioinformatics is marked by the development of an unparalleled number of software and computational tools, devoted to dealing with massive and ever-increasing amounts of data, with the number now doubling in under two years[41]. Handling, processing and storing information have become new challenges for biologists, embodied in a single concept - big data, a term to describe the challenge of dealing with several terabytes of data. An additional challenge that the newly developed tools are facing is how to extract, interpret and conceptualize the inherently complex cellular systems and processes from voluminous, yet relatively simple, collected datasets. To help biologists navigate in this challenging environment, a number of initiatives have emerged with the goal to classify, index, and share a maximum of available tools[7,42–44].

# Evolution of bioinformatics tools

While the origins of bioinformatics can be easily retraced and reviewed, it is much more difficult to consider its evolution over the past decade, mainly because the key messages are lost in the wealth of repositories, databases, and publications available online. Who are the main providers of bioinformatics tools? What are the promising and trending fields in bioinformatics, and which areas are losing interest? To answer these questions, we used the freely accessible OMICtools database[7] to extract a compilation of data on bioinformatics tools with the goal to provide a perspective on the evolution of modern bioinformatics tool development.

Among all currently available bioinformatics tools repositories, OMICtools indexes the largest number of entries, and provides a comprehensive, updated ontology-based classification system that tracks various pertinent parameters and metadata for each tool. This includes its technology and/or analysis step, the date and journal of publication, country and institute of development, coding language, availability, past and current versions and usage scenario[7,45]. The repository includes bioinformatics tools identified and extracted from scientific publications and software repositories using automated mining algorithms which were subsequently manually curated and categorized. We analyzed data from over ~23,000 tools developed between 1990 and 2017. This comprehensive dataset enables us to analyze the bioinformatics tools landscape in various contexts, including the growth of available tools, their popularity or their lifespan throughout the years (Supplementary Figure 1). For the sake of simplicity, in this review, we apply the term "tool" for either a software tool or a database resource. Since the 2000s and the completion of the Human Genome Project, the number of tools being developed has grown exponentially, today doubling in under four years (Figure 1a); to put this in perspective, five times more tools were published in the last year than the total number of tools published between 1990 and 2000. This trend clearly brings to light an increasing challenge for researchers: finding the right tool for optimal data analysis in a constantly changing field.

**Tools applications and technologies**

In the OMICtools database, bioinformatics tools are classified according to their scope, in one or several of the following categories: genomics, transcriptomics, proteomics, metabolomics, epigenomics or phenomics. Figure 1b presents the number of tools produced each year since 1990 according to omics application, demonstrating that all of the primary omics fields have seen a constant increase in the annual production of tools since the early 90s.

To follow the dynamics of omics technologies, the annual production of bioinformatics tools dedicated to well-known applications in HTS, microarrays, PCR, mass-spectrometry, nuclear-magnetic resonance (NMR), flow-cytometry (FC) and bioimaging are presented in Figure 1c. New technologies in HTS, such as RNA-sequencing and single-cell RNA-sequencing, generated a rapid and exponential increase in the number of dedicated tools, while the rate of tool production in microarrays or Sanger sequencing, that predate new high-throughput technologies, are plateauing. Since a given tool can be used for more than one technology or analysis step, we extracted the number of duplications for each tool, that is the number of applications, analysis step, or function associated with a tool. The vast majority of tools in our database (82%) are assigned to one application, while 18% of tools are assigned to two or more applications (Supplementary Figure 2).

**Tool development worldwide**

Tool development and publication is dominated by the USA, with 9841 out of 30,141 (32,6%)

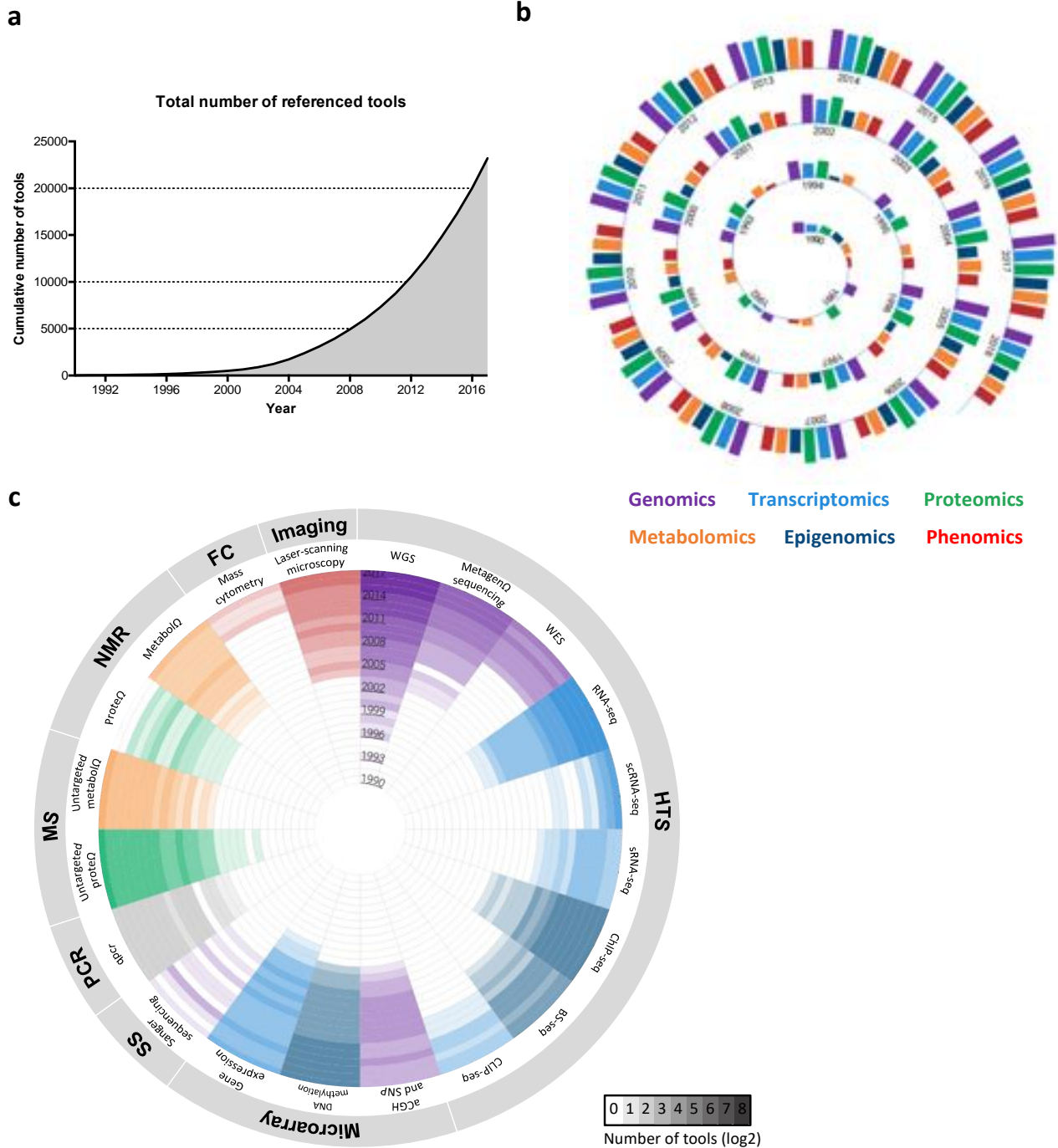

**Figure 1: Evolution of bioinformatics tools from 1990 to 2017.**
(a) Cumulative production of tools among 23,201 tools classified in the OMICtools repository. (b) Evolution of the number of tools produced annually according to main omics field. *(c)* Number of tools published annually from 1990 to 2017 by main omics technology. Sub-technologies that belong to a specific omics fields are color-coded. qPCR tools can be either classified as genomics or transcriptomics technology and are therefore represented in gray. HTS: High-throughput sequencing; MS: Mass spectrometry; NMR: Nuclear-magnetic resonance; FC: Flow-cytometry; WGS: Whole-genome sequencing; WES: Whole-exome sequencing; SS: Sanger sequencing; Ω: Omics. Data were transformed in Log2 for better clarity.

affiliations on published tools being US institutions (Figure 2a and b, full list of countries provided in Table 1). Most leading tool-developing institutions are hosted by European and American continents, with 18 out of the top 20 tool-developing institutes and universities located in the USA, United Kingdom, Canada, or rest of Europe (Supplementary Figure 3).

Interestingly, supplementing the specialized bioinformatics-centered institutions, institutes and universities leading in life sciences also have ample in-house capacity to produce their own tools and resources. Indeed, bioinformaticians are highly sought after profiles in every biology lab[46].

Strikingly, no African countries appear in the top 30 tool-developing countries (Table 1), although this might change with the new H3Africa initiative begun in 2010 and other national initiatives[47,48]. Despite awareness and progressive efforts to increase accessibility of knowledge of bioinformatics and computing skills, tool development remains largely the domain of countries with advanced scientific and financial resources[49], even though open access and open data have created opportunities for those with capable internet access and relatively modest compute facilities. Figure 2c shows a high correlation (r=0.81; p<0.0001) between a country's research and development (R&D) expenditure in GDP and the number of tools published by one of its institutions.

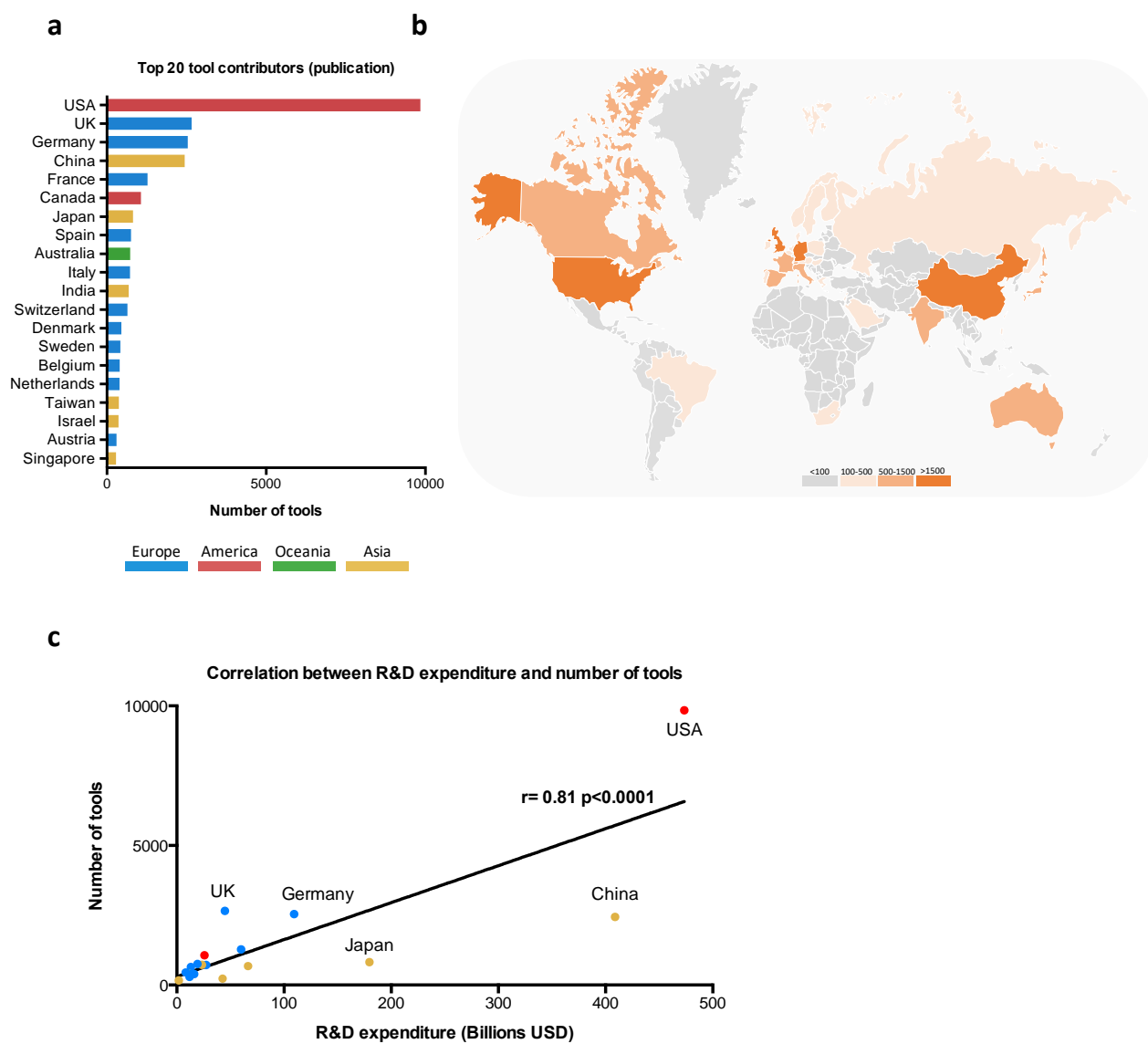

**Figure 2: Tool development worldwide.**
(a) Number of tools published per countries (top 20). (b) Color-coded world map representation of the number of tools produced per country. (c) Correlation between research and development expenditure (UNESCO data) and total number of tools produced by the top 20 countries with highest Gross Domestic Product (GDP).

Scientific research relies heavily on collaborations between international institutions. Similarly, publications of bioinformatics tools often result from collaborations. In our database, a total of 5063 published tools are affiliated with more than one unique country (Figure 3a). This translates into approximately 22% (5063 out of 22,891 tools for which information was retrieved) of tools developed through an international collaboration. Among those, figure 3b represents the network formed by the top 40 collaborative pairs of countries. The USA interacts widely, participating in the highest number of collaborations, distributed between the UK (17.3%), China (16.8%), Germany (12.3%), and the rest of Europe and Asia (Figure 3c). In contrast, the USA represents the vast majority (74%) of China's collaborations for tool development (Figure 3d). This analysis also highlights interesting parameters that potentially play a role in promoting collaboration between countries, such as geographical proximity (Australia and China) and common spoken languages (France and Switzerland, Austria and Germany).

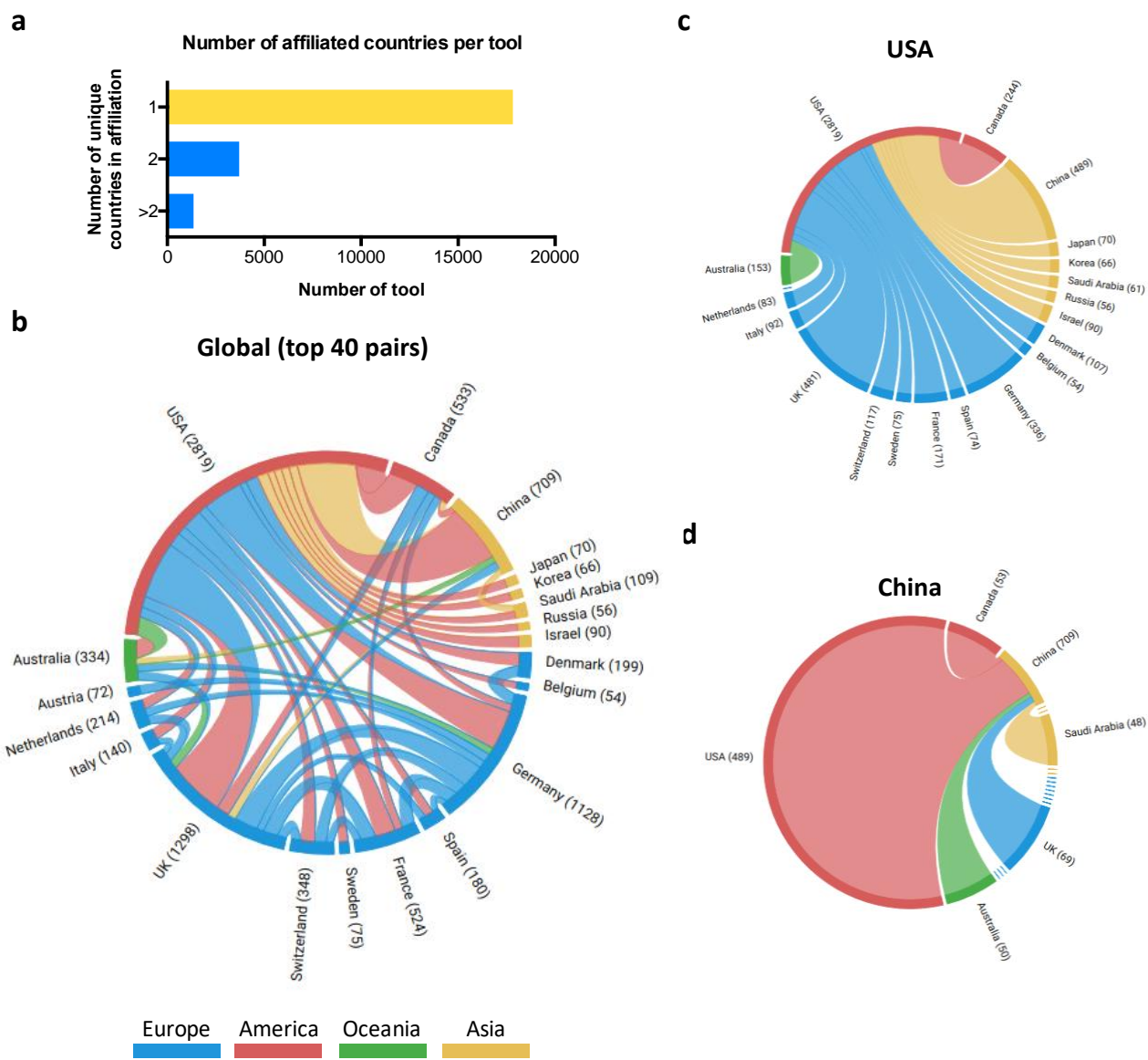

**Figure 3: Collaborations for tool development.**
(a) Number of unique affiliated countries per tool publication, among a total of 22,891 tools. (b-d) Network formed by the top 40 paired collaborations for tool publications. The total number of collaborations are shown for all countries (b) and between paired countries (c and d). If a tool publication has more than 2 affiliated countries, each pair is counted. An interactive version of these collaborations is available at https://omictools.com/bioinformatics-trends#chord-graph.

**Economics of tool development**

Funding is a critical aspect of tool development. We ranked the sources of funding agencies associated with a total of 12,761 published tools to assess the top 20 tool-funding agencies worldwide (Figure 4). Nearly half of all published tools were funded by the National Institutes of Health (NIH) or the National Science Foundation (NSF), both US-based agencies. Over the past few years (and arguably since its foundation), the field of bioinformatics has developed in the direction of both open-source data and software. Freely available tools and frameworks have already served as a foundation for building important applications and resources[50]. In the OMICtools database, 270 out of 7,838 (3.4%) tools with a known license are registered with a commercial license while the remaining 7,568 tools are open-source (data not shown), confirming the commitment of bioinformatics tools developers to support open science.

**Bioinformatics tools in the literature**

Nearly eight out of ten tools in the database have been published in a peer-reviewed journal (Figure 5a), with 41.7% of all tools cited at least once, while 36.0% have never been cited in the literature. We analyzed the extent of tool citation in relation to the timing of development of a new technology, taking into account their publication age. (Figure 5b). A clear trend emerged as seen with the example of RNA-sequencing technology, with the first papers published in 2008; tools dedicated to analysis of RNA-sequencing data that were published in 2009 are on average significantly more cited that tools published in subsequent years ($P < 0.05$). This trend was observed with other technologies including WGS, CHIP-seq and CLIP-seq, suggesting that tools that are the first to resolve a problem are more likely to be established as gold standards or default methods, and by consequence accumulate more citations over time.

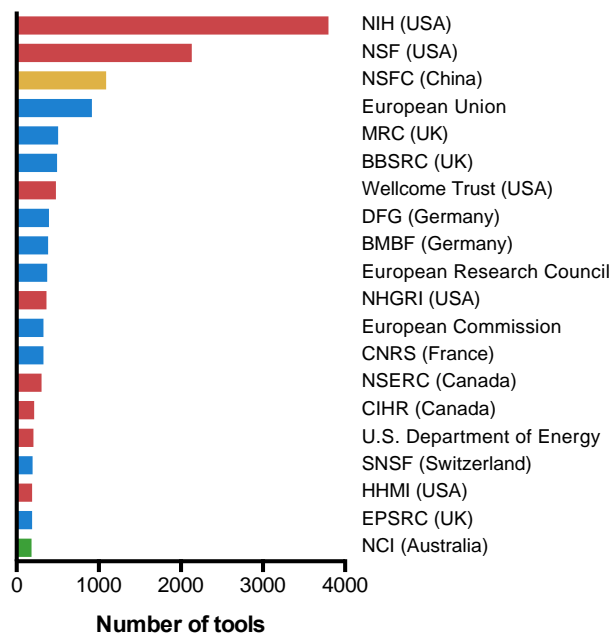

**Figure 4: Top funding agencies.**
Number of published tools by funding institutions (top 20).

**Tool specifications**

The user can run bioinformatics software tools either on the web, locally on a desktop or server, or both. While tools that can be used on the web could be expected to be more common, reflecting the need for user-friendliness for less-skilled users, we in fact found that more than 69% of the 20,918 tools registered as software are developed as desktop applications only (Figure 6a). Similarly, we speculated that the majority of tools could be run on the mainstream operating systems (OS) Windows and Mac OS; we in fact found that more than half of the 15,736 tools with a known OS can be run on UNIX/Linux exclusively (Figure 6b). Moreover, 42.6% of tools are usable on more than one OS (34.2% on all three). Figure 6c represents the number of tools produced each year by the top-6 most used programming languages, plus Fortran and Javascript. Interestingly, while Java was the most used programming language each year from 2000 to 2010, it has now been surpassed by R, Python, and C++.

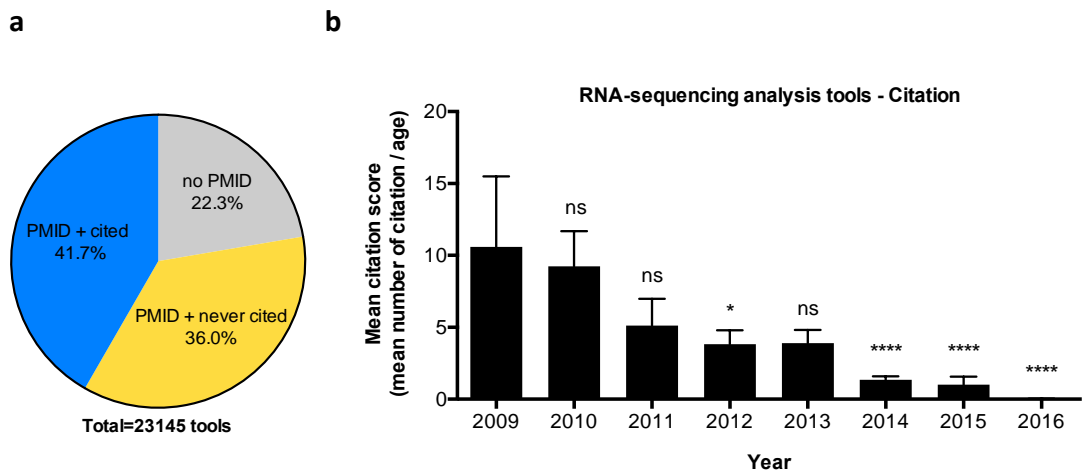

**Figure 5: Bioinformatic tools in the literature.**
(a) Proportion of tools in the OMICtools database that are not associated with a PMID (gray), that have a PMID but have never been cited (yellow), or that have been cited at least once (blue). (b) Mean citation score of all tools for RNA-sequencing analysis by publication year. For each tool with an associated PMID, the citation score is the number of time this tool has been cited in the literature divided by its publication age (the number of year since its publication; 2017 - year of publication). The bar plot represents the mean citation score for all tools published in a given year + standard error of mean (SEM). Statistical significance is indicated by $*p < 0.05$, $****p < 0.0001$. Kruskal-Wallis multiple comparison test was used to compare every group to the "2009" group. ns: not significant.

Overall, these results likely reflect the fact that software tools are usually developed to address a need, pipeline, or problem that is specific to their developer, and are not necessarily designed to be easy-to-use for the average biologist. These data also indicate that tool development remains a field that requires specific sets of skills and highlights the importance of interactions between the biologists who use them and the bioinformaticians creating them.

**From individual tools to pipelines**

For an arguably extended period, biological data consisted of a handful of sequences to be analyzed and compared, which could be done in a few computational steps and by using a single program. However, due to their complexity and quantity, to obtain meaningful data, today's biological datasets require multiple analysis steps that often need a series of different programs that must be run in a specific order. To verify this, we followed the evolution of co-citations (the number of tools cited per publication) and observed an increase in the number of tools jointly cited in scientific publications over time (Figure 7). While publications in the early 2000s were citing one to five tools at most, the number of tools cited per publication has continuously increased since 2005, with 20% of publications in 2015 citing more than six tools. This indicates a shift in biological data complexity, now requiring the use of pipelines of tools for effective and productive analysis.

## Conclusion

The present review proposes an illustrated and data-supported historical perspective of the evolution of bioinformatics since the 1990s. By collecting, sorting, and analyzing freely available data from more than ~23,000 bioinformatics tools, we provide a collection of original insights into the subtleties of this field. After years of parallel development, biology and informatics joined forces in the early 60s to rapidly become a major field in the biological sciences, dedicated to the handling and analysis of omics data. The last decade has seen a rapid expansion in the number of bioinformatics tools. Initially developed by biologists in need to respond to a specific question and accelerate their data analyses, they are now complex pieces of software that require advanced programming and coding knowledge. They necessitate a growing amount of time investment and financial resources, they are increasingly developed by extensive collaborative networks, and are published in high-quality peer-reviewed journals. The acknowledged scientific value of bioinformatics tools will only increase and expand to include other fields in the future. Here we highlight some insights from developments since the inception of the field, which shed light on the short-term direction and optimization of data analysis using bioinformatics tools.

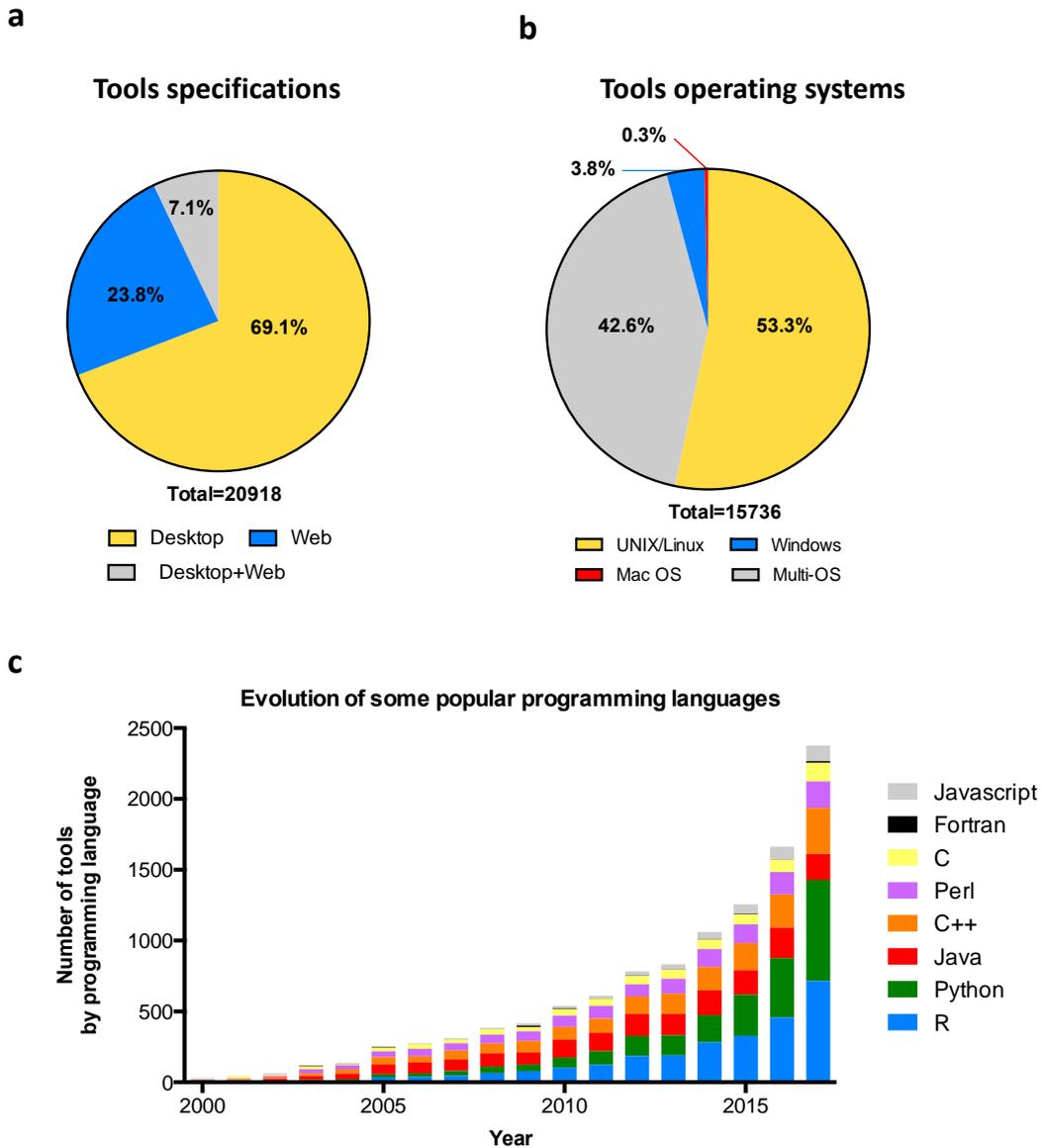

**Figure 6: Programming language and tool specification.**
(a) Proportion of tools by interface, among all tools except datasets (n=20918). (b) Proportion of tools by operating systems (n=15736). (c) Number of tools produced each year by programming languages. Only the top 7 most-used programming languages and Fortran are represented.

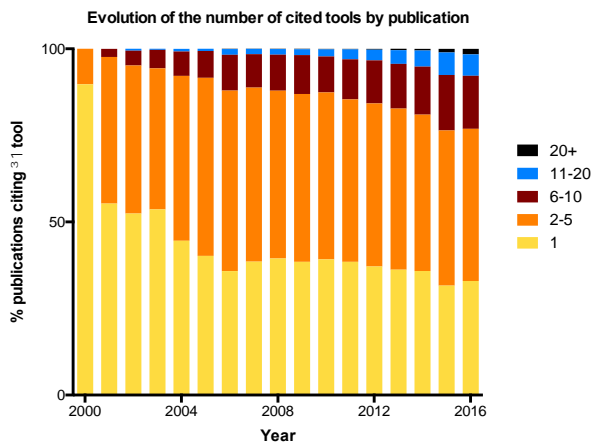

**Figure 7: Tool pipelines in the literature.**
Evolution of the proportion of tools cited per publication among publications citing at least one tool from 2000 to 2015. Individual publications citing at least one tool registered in the OMICtools database were retrieved using the MEDLINE API.

Paradoxically, while bioinformatics tools were designed to deal with "big data", their proliferation now raises concerns about a possible "big tool" pool, representing several resources so great that researchers primary challenge will be locating and choosing the right tools in order to explore their data. While databases and repositories propose the indexation of a considerable number of tools, they have trouble keeping pace with their constantly growing publication rate. As our data analyses suggest, analysis of complex biological datasets increasingly requires software to be run in pipelines of tools. Future challenges for biologists will thus include rapid identification of the most appropriate and functional combinations of tools, and the execution of these pipelines in a fully-automated manner, with a minimum of resources and time, requiring the increasing involvement of machine learning and artificial intelligence methods, not only in data analysis, but also in tool selection.

## Methods

On December 8th 2017, a total of 23,201 tool entries were extracted from the OMICtools database (https://omictools.com). The following data were retrieved (if available): date of creation, ontology (OMICtools classification), journal of publication, number of citations, countries and institutions associated with authors, source of funding, programming language and specifications, and the URL of the tool. To draw a correlation between countries' R&D expenditure and number of tools produced, data from UNESCO (gross domestic expenditure on R&D, available at http://data.uis.unesco.org/Index.aspx and accessed in January 2018) for the top-20 countries with the highest GDP, were used.

To evaluate the impact of publication year on mean citation number, we calculated a mean "citation score" for all RNA-sequencing analysis tools by publication year. To account for the fact that older articles tend to accumulate more citations, we divided the citation count for each tool with an associated PMID by its publication age, calculated as 2017 minus the year of publication, and expressed mean citation scores by years. Tool duplications within the RNA-sequencing category were removed, and tools with zero citations were kept in the analysis. A Kruskal-Wallis multiple comparison test was used to compare every group to the "2009" group.

Some figures of this article are dynamic and available at https://omictools.com/bioinformatics-trends.

## Acknowledgements

The authors are grateful to the additional omicX employees who contributed to the manuscript. We thank Stephen Altschul for critically reviewing this manuscript, Sarah Mackenzie for English proofreading of the manuscript, Marion Denorme and omicX founder Arnaud Desfeux for discussion and advice. KV is supported by the European Structural and Investment Funds grant for the Croatian National Centre of Research Excellence in Personalized Healthcare (contract #KK.01.1.1.01.0010), Croatian National Centre of Research Excellence for Data Science and Advanced Cooperative Systems (contract KK.01.1.1.01.0009 and Croatian Science Foundation (grant IP-2014-09-6400). DL was supported by the Intramural Research Program of the National Institutes of Health, National Library of Medicine.

## Author Contributions

C.L. analyzed data; drafted and finalized the manuscript, E.D. extracted and analyzed data, B.G.; D.L.; E.H.; L.M. contributed to manuscript preparation/writing/reviewing, K.V. contributed to project management and reviewed/edited the manuscript.

## Competing Interests Statement

The authors declare the following competing interests: CL and ED are employed at omicX, the provider of OMICtools.

## References

1. Edman, P. & Begg, G. A Protein Sequenator. *Eur. J. Biochem.* 80–91 (1967).

2. Needleman, S. B. & Wunsch, C. D. A general method applicable to the search for similarities in the amino acid sequence of two proteins. *J. Mol. Biol.* **48,** 443–453 (1970).

3. Mortazavi, A., Williams, B. A., McCue, K., Schaeffer, L. & Wold, B. Mapping and quantifying mammalian transcriptomes by RNA-Seq. *Nat. Methods* **5,** 621–628 (2008).

4. Kallio, M. A. *et al.* Chipster: user-friendly analysis software for microarray and other high-throughput data. *BMC Genomics* **12,** (2011).

5. Stephens, Z. D. *et al.* Big Data: Astronomical or Genomical? *PLOS Biol.* **13,** e1002195 (2015).

6. Karsch-Mizrachi, I., Takagi, T., Cochrane, G. & on behalf of the International Nucleotide Sequence Database Collaboration. The international nucleotide sequence database collaboration. *Nucleic Acids Res.* **46,** D48–D51 (2018).

7. Henry, V. J., Bandrowski, A. E., Pepin, A.-S., Gonzalez, B. J. & Desfeux, A. OMICtools: an informative directory for multi-omic data analysis. *Database* **2014,** bau069–bau069 (2014).

8. Hagen, J. B. The origins of bioinformatics. *Nat. Rev. Genet.* **1,** 227–231 (2000).

9. Hogeweg, P. The Roots of Bioinformatics in Theoretical Biology. *PLoS Comput. Biol.* **7,** e1002021 (2011).

10. Watson, J. D. & Crick, F. H. Molecular structure of nucleic acids. *Nature* **171,** 737–738 (1953).

11. Gerola, H. & Gomory, R. E. Computers in science and technology: Early indications. *Science* **225,** 11–18 (1984).

12. Kendrew, J. C., Bodo, G., Dintzis, H. M., Parrish, R. G. & Wyckoff. A three-dimensional model of the myoglobin molecule obtained by X-ray analysis. *Nature* **181,** 662–666 (1958).

13. Simpson, H. R. The estimation of linkage on an electronic computer. *Ann. Hum. Genet.* **22,** 356–361 (1958).

14. Sanger, F. & Tuppy, H. The amino-acid sequence in the phenylalanyl chain of insulin. 1. The identification of lower peptides from partial hydrolysates. *Biochem. J.* **49,** 463 (1951).

15. Dayhoff, M. O. & Richard, E. V. Atlas of Protein Sequence and Structure: 1967-68. in (1969).

16. Dayhoff, M. O. Computer aids to protein sequence determination. *J. Theor. Biol.* **8,** 97–112 (1965).

17. Dayhoff, M. O. & Ledley, R. S. Comprotein: a computer program to aid primary protein

structure determination. *Proc. Dec. 4-6 1962 Fall Jt. Comput. Conf.* 262–274 (1962).

18. Doolittle, R. F. Amino-acid sequence investigations of fibrinopeptides from various mammals: Evolutionary implications. *Nature* **202,** 147–152 (1964).

19. Fitch, W. M. & Margoliash, E. Construction of Phylogenetic Trees. *Science* **155,** 279–284 (1967).

20. Holley, R. *et al.* Structure of a Ribonucleic Acid. *Science* **147,** 1462–1465 (1965).

21. Sanger, F. Sequences, sequences, and sequences. *Annu. Rev. Biochem.* **57,** 1–29 (1988).

22. Gibbs, A. J. & McIntyre, G. A. The diagram, a method for comparing sequences. *FEBS J.* **16,** 1–11 (1970).

23. Tinoco, I. J., Uhlenbeck, O. C. & Levine, M. D. Estimation of Secondary Structure in Ribonucleic Acids. *Nature* **230,** 362–367 (1971).

24. Jordan, B. R. Computer generation of pairing schemes for RNA molecules. *J. Theor. Biol.* **34,** 363–378 (1972).

25. Zhurkin, V. B., Lysov, Y. P. & Ivanov, V. I. Computer analysis of conformational possibilities of double-helical DNA. *FEBS Lett.* **59,** 44–47 (1975).

26. Staden, R. Sequence data handling by computer. *Nucleic Acids Res.* **4,** 4037–4052 (1977).

27. Maxam, A. M. & Gilbert, W. A new method for sequencing DNA. *Proc. Natl. Acad. Sci. U. S. A.* **74,** 560–564 (1977).

28. Sanger, F., Nicklen, S. & Coulson, A. R. DNA sequencing with chain-terminating inhibitors. *Proc. Natl. Acad. Sci. U. S. A.* **74,** 5463–5467 (1977).

29. Glowniak, J. History, structure, and function of the internet. *Telenuclear Med.* **28,** 135–144 (1998).

30. Queen, C. & Korn, L. J. A comprehensive sequence analysis program for the IBM personal computer. *Nucleic Acids Res.* **12,** 581–599 (1984).

31. Bilofsky, H. S. & Christian, B. The GenBank® genetic sequence data bank. *Nucleic Acids Res.* **16,** 1861–1863 (1988).

32. Hamm, G. H. & Cameron, G. N. The EMBL data library. *Nucleic Acids Res.* **14,** 5–9 (1986).

33. Altschul, S. F., Gish, W., Miller, W., Myers, E. W. & Lipman, D. J. Basic local alignment search tool. *J. Mol. Biol.* **215,** 403–410 (1990).

34. Pearson, W. R. & Lipman, D. J. Improved tools for biological sequence comparison. *Proc. Natl. Acad. Sci.* **85,** 2444–2448 (1988).

35. Fleischmann, R. *et al.* Whole-genome random sequencing and assembly of Haemophilus influenzae Rd. *Science* **269,** 496 (1995).

36. Goffeau, A. *et al.* Life with 6000 Genes. *Science* **274,** 546 (1996).

37. International Human Genome Sequencing Consortium. Initial sequencing and analysis of the human genome. *Nature* **409,** 860 (2001).

38. Venter, J. C. *et al.* The Sequence of the Human Genome. *Science* **291,** 1304 (2001).

39. Likić, V. A., McConville, M. J., Lithgow, T. & Bacic, A. Systems Biology: The Next Frontier for Bioinformatics. *Adv. Bioinforma.* **2010,** 1–10 (2010).

40. Reuter, J. A., Spacek, D. V. & Snyder, M. P. High-Throughput Sequencing Technologies. *Mol. Cell* **58,** 586–597 (2015).

41. Marx, V. Biology: The big challenges of big data. *Nature* **498,** 255–260 (2013).

42. Gentleman, R. C. *et al.* Bioconductor: open software development for computational biology and bioinformatics. *Genome Biol.* **5,** R80 (2004).

43. Gnimpieba, E. Z., VanDiermen, M. S., Gustafson, S. M., Conn, B. & Lushbough, C. M. Bio-TDS: bioscience query tool discovery system. *Nucleic Acids Res.* **45,** D1117–D1122 (2017).

44. Ison, J. *et al.* Tools and data services registry: a community effort to document

bioinformatics resources. *Nucleic Acids Res.* **44,** D38–D47 (2016).

45. Perrin, H. *et al.* OMICtools: a community-driven search engine for biological data analysis. *arXiv:1707.03659* (2017).

46. Chang, J. Reward bioinformaticians. *Nature* **520,** (2015).

47. Karikari, T. K. Bioinformatics in Africa: The Rise of Ghana? *PLOS Comput. Biol.* **11,** e1004308 (2015).

48. The H3Africa Consortium. Enabling the genomic revolution in Africa. *Science* **344,** 1346 (2014).

49. Pongor, S. & Landsman, D. Bioinformatics and the developing world. *Biotechnol. Dev. Monit.* **40,** 10 (1999).

50. Stajich, J. E. Open source tools and toolkits for bioinformatics: significance, and where are we? *Brief. Bioinform.* **7,** 287–296 (2006).

51. Callahan, A., Winnenburg, R. & Shah, N. H. U-Index, a dataset and an impact metric for informatics tools and databases. *Sci. Data* **5,** 180043 (2018).14